\documentclass[journal]{IEEEtran}

\usepackage{algorithmic}
\usepackage{graphicx}
\usepackage{array}  
\usepackage[cmex10]{amsmath}
\usepackage{amsfonts}
\usepackage{umoline}
\usepackage{color}
\usepackage{cite}

\DeclareMathOperator{\SINR}{SINR}

\allowdisplaybreaks[1]
\begin{document}

\title{Unified and Distributed QoS-Driven Cell Association Algorithms in Heterogeneous Networks}

\author{\IEEEauthorblockN{H. Boostanimehr,  and V. K. Bhargava, }
\thanks{H. Boostanimehr and V. K. Bhargava are with Department of Electrical and Computer Engineering, University of British Columbia, Vancouver, Canada (E-mail: \{hamidrbm, vijayb\}@ece.ubc.ca).}
%\thanks{Parts of this work have been submitted to 2014 Iran Workshop on Communication and Information Theory (IWCIT'14) for possible publication.}
\thanks{This work is funded in parts by the Natural Sciences and Engineering Research Council of Canada under their Strategic Project Grants Program.}

%\author{\IEEEauthorblockN{Hamidreza Boostanimehr, \emph{Member, IEEE} and Vijay K. Bhargava, \emph{Fellow, IEEE}}
%\thanks{H. Boostanimehr and V. K. Bhargava are with Department of Electrical and Computer Engineering, University of British Columbia, Vancouver, Canada (E-mail: \{hamidrbm, vijayb\}@ece.ubc.ca).}
%\thanks{This work is funded in parts by the National Sciences and Engineering Research Council of Canada under their Strategic Project Grants Program.}

} 
%\date{today}
 
\maketitle

\begin{abstract}
This paper addresses the cell association problem in the downlink of a multi-tier heterogeneous network (HetNet), where base stations (BSs) have finite number of resource blocks (RBs) available to distribute among their associated users. Two problems are defined and treated in this paper: sum utility of long term rate maximization with long term  rate quality of service (QoS) constraints, and global outage probability minimization with outage QoS constraints. The first problem is well-suited for low mobility environments, while the second problem provides a framework to deal with environments with fast fading. The defined optimization problems in this paper are solved in two phases: cell association phase followed by the optional RB distribution phase. We show that the cell association phase of both problems have the same structure. Based on this similarity, we propose a unified distributed algorithm with low levels of message passing to for the cell association phase. This distributed algorithm is derived by relaxing the association constraints and using Lagrange dual decomposition method. In the RB distribution phase, the remaining RBs after the cell association phase are distributed among the users. Simulation results show the superiority of our distributed cell association scheme compared to schemes that are based on maximum signal to interference plus noise ratio (SINR).
\end{abstract}


\section{Introduction}
\PARstart{C}{onventional} cellular networks are homogeneous networks composed of similar base stations (BSs) which are carefully planned in a given geographical area. The similarity of the BSs in homogeneous networks is in having high transmit power and the number of users they can support. Coverage holes in conventional networks are expected because of the random behaviour of wireless channels in urban and rural areas. Moreover, the number of mobile subscribers and mobile data demand have been showing an unprecedented growth in recent years \cite{hasan2011green}. In order to cope with this explosive growth, 3GPP LTE has been studying heterogeneous networks (HetNet) which are cellular in nature, and increase spectral efficiency per unit area \cite{Madan-1,Survey_on_HetNet-1, LTE1}. HetNets are composed of macro BSs overlaid with lower tier BSs (BSs with lower powers) such as pico, femto, and relay  BSs. Macro BSs with their high transmit power cover large geographical areas, while randomly scattered lower tier BSs serve users in coverage holes and hot spots.

HetNets, with their increased diversity in type and number of BSs, re-open the conventional challenges in cellular wireless networks. Among these challenges are cell association, resource allocation, and intercell interference management (ICIC) \cite{Survey_on_HetNet-1}. Cell association rules, which are the focus of this paper, are a set of rules that determine which BS serves a particular user. The cell association rule in conventional cellular networks, and up to LTE release-8, has been based on the strongest signal to interference plus noise ratio (SINR) seen by the user; each user associates itself with the BS that provides the best SINR \cite{3gpp.36.912}. Following this rule in HetNets may deviate the performance from optimality because macro BSs have higher power and attract more users than low tier BSs. Since the resources available at a BS are limited, more users associated with a BS translates into less resources available for a given user, which in turn reduces the throughput seen by that user. Although the users associated with overloaded BSs are experiencing high levels of SINR, their throughput is reduced. Therefore, achieving load balancing among BSs in cell association phase becomes as important as providing high levels of SINR. Cell association problem, although well-studied for conventional networks, has not been thoroughly addressed in the context of HetNets.

Formulating the cell association problem naturally falls into the scope of integer or mixed integer programming since each user is to be mapped to a BS. There are several approaches to cope with these integer programs to achieve load balancing in the cell association phase. Solving the integer program directly by exploiting the structure of the problem \cite{Direct1, Direct2}, relaxing the association constraints and using Lagrange dual decomposition method \cite{Andrews-association,Andrews2, Relax1}, Markov decision process frameworks \cite{MDP1, MDP2}, game theoretic frameworks \cite{Game1,Game2}, and stochastic geometry frameworks \cite{GeoSto_Model2, GeoSto_Model3, GeoSto_Model4, GeoSto1} are examples of these approaches among others. In the next two paragraphs, we focus on the first two approaches as they are more relevant to the work presented in this paper. 

Authors of \cite{Direct1} focus on flow level cell load balancing and association under spatially inhomogeneous traffic distributions. A unified distributed and iterative algorithm is proposed that adapts to traffic loads and converges to the optimal point. The objective function of the defined optimization problem in \cite{Direct1} can be selected from a family of objective functions; each of which directs the solution towards a rate, throughput, delay, or load balance optimal point. In \cite{Direct2}, an online algorithm is developed based on the idea of associating users with BSs that provide the best expected throughput instead of associating users with the best SINR providing BS. Considerable interference avoidance and load balancing gains are achieved through the online algorithm proposed in \cite{Direct2}. In these two works, the association constraints are not relaxed and the proposed algorithms produce binary decision variables.

In many cases, formulating the cell association problem leads to NP-hard assignment problems. Relaxing the association constraints and applying Lagrange dual decomposition method is a popular method to cope with these cases. This is because relaxing the association constraints usually converts the optimization problem into a convex or linear program, for which efficient algorithms exist. Additionally, dual decomposition methods usually lead to distributed algorithms, where the nodes in the network decide based on their local information, as opposed to centralized solutions which require global information access at one central node or all the nodes. Examples of distributed cell association algorithms can be found in \cite{Andrews-association,Andrews2, Relax1}. In all these works, the resources at each BS is distributed evenly among the users associated with that BS. In \cite{Andrews-association}, with which our paper has the most correlation in terms of system model and solution approach, it is proven that distributing the resources equally among the users connected to a given BS is optimal for a logarithmic objective function. Based on this observation, a distributed algorithm is proposed that converges to a near optimal point to improve the long term rate.   Range expansion technique by biasing the SINR of lightly loaded BSs to make them more attractive to the users is also incorporated in the distributed algorithm in \cite{Andrews-association}. It is worth noting that cell range expansion through cell biasing is a simple and effective load balancing scheme that is being discussed by the 3GPP in the context of LTE-Advanced \cite{Survey_on_HetNet-1}. Extending over \cite{Andrews-association}, in \cite{Andrews2}, the joint optimization of load balancing and enhanced intercell interference coordination (referred to as eICIC by 3GPP) via almost blank subframes (ABS) is considered. Lastly, \cite{Relax1} is another work in which a dynamic cell association and cell range expansion algorithm for load balancing through relaxation of association constraints is proposed. The algorithm presented in \cite{Relax1} is in a centralized fashion. 

In recent standards such as LTE, the resources at BSs are distributed among users in the form of resource blocks (RBs). An RB spans over a certain frequency range and time duration. Depending on the total channel bandwidth available at a BS and the scheduling interval of the scheduler, the number of RBs at different BSs can be different. The RB-based structure in standards such as LTE-Advanced results in more flexible resource allocation schemes, thus higher spectral efficiency \cite{cheng2011exploiting}. RBs are assigned to users to satisfy their quality of service (QoS) requirements. Most of the works on cell association problem in the literature, including the aforementioned works in this section,  have not considered the RB-based structure of the BS resource budget in their system models. In an attempt to follow the LTE system model, in this paper, we consider finite number of RBs at each BS as its resource budget.

The accomplished works in this paper are briefly described here. This paper addresses the cell association problem in the downlink of a multi-tier HetNet, where BSs have finite number of RBs available to distribute among their associated users. We investigate distributed algorithms where users and BSs decide based on their local measurements of the wireless environment. We also focus on providing QoS in terms of minimum achievable long term rate or maximum outage probability. Two problems are defined and treated in this paper: sum utility of long term rate maximization with long term rate QoS constraints, and global outage probability minimization with outage QoS constraints. The first problem is well-suited for low mobility environments, while the second problem provides a framework to deal with environments with fast fading. In defining the optimization problems in this paper, we consider a general scenario where frequency reuse factor of $1$ and no interference coordination schemes are assumed. Both problems are to be optimized through cell and RB association. The defined optimization problems in this paper are solved in two phases: cell association phase followed by the optional RB distribution phase. We show that the cell association phase for both problems have the same structure. Based on this similar structure, we design and propose a unified distributed algorithm with low levels of message passing and complexity for the cell association phase. The distributed cell association algorithm is derived by relaxing the association constraints and using Lagrange dual decomposition method. Our distributed cell association algorithm is QoS-driven since users receive only enough number of RBs to satisfy their QoS constraints while maximizing the sum utility of rate or minimizing the global outage probability. In the RB distribution phase, the remaining RBs after the cell association phase are distributed among the users for further improvement of the network performance. The RB distribution phase for the rate problem is a convex program with a closed form solution, while distributing the remaining RBs for the outage problem is a complex non-convex non-linear problem, for which we propose a sub-optimal greedy algorithm. Extensive simulation results that are brought in this paper show that our distributed cell association scheme outperforms the maximum SINR scheme. For instance, rate gains of up to 2.4x have been observed in the simulations for the cell edge users in our distributed cell association algorithm over maximum SINR scheme. 

The rest of the paper is organized as follows: In the next section, the chosen system model is described. The cell association problem with QoS constraints is formulated in section III. Our distributed solution to the cell association problem is presented in section IV. Section V will address distributing the remaining RBs after the cell association phase. In section VI, we examine the performance of our proposed algorithms through numerical simulations, and finally, section VII concludes the paper.

%%%%%%%%%%%%%%%%%%%%%%%%%%%%%%%%%%%%%%%%%%%%%%%%%%%%%%%%%%%%%%%%%%%%%%%%%%%%%%%%%%%
%%%%%%%%%%%%%%%%%%%%%%%%%%%%%%%%%%%%%%%%%%%%%%%%%%%%%%%%%%%%%%%%%%%%%%%%%%%%%%%%%%%
%%%%%%%%%%%%%%%%%%%%%%%%%%%%%%%%%%%%%%%%%%%%%%%%%%%%%%%%%%%%%%%%%%%%%%%%%%%%%%%%%%%
\section{System model}
\label{sec:sys_model}
The focus of this paper is on a downlink HetNet consisting of multiple tiers of BSs, where different tiers represent different types of BSs. As an example, tier 1 BSs can be macro BSs with high transmit power and large coverage areas. Tier 2 and 3 BSs (pico and femto BSs) are regarded as smaller BSs with lower powers compared to the tier 1 BSs, but with higher deployment density. Finally, tier 4 BSs model indoor access points with very small transmit powers.

%\begin{figure}[tb]
%\label{fig:sys_model}
%\centering
%\includegraphics[width=3.5in]{HetNet_Sys_Model.pdf}
%\caption{A $K$-tier HetNet downlink with $K=4$}
%\end{figure}

The set of all BSs is denoted by $\mathcal{B}=\{1, \dots, N_\mathcal{B}\}$, and the set of all users is denoted by $\mathcal{U}=\{1, \dots, N_\mathcal{U}\}$. The cardinality of $\mathcal{B}$ is $N_\mathcal{B}$, and the cardinality of $\mathcal{U}$ is $N_\mathcal{U}$. Each BS $j \in \mathcal{B}$ has a fixed power of $P_j$ Watts available. All the BSs are assumed to be connected by a high speed backhaul through which information exchange with negligible delay is possible.

\subsection{Notion of Resource Blocks}
%One of the main contributions of this paper is to use the notion of resource blocks (RB) in our system model. The rational behind this is to get as close as possible to system models defined in LTE, which is a very promising technology today. In this subsection, we elaborate the concept of RBs.

In recent years, data rate hungry applications have derived the wireless network researchers and vendors to develop OFDMA-based LTE-Advanced networks. OFDMA technology makes possible a flexible resource structure, where the time-frequency spectrum is divided into orthogonal resource blocks (RBs). The RB-based structure in LTE-Advanced  allows for more flexible resource allocation schemes, thus results in higher spectral efficiency \cite{cheng2011exploiting}.   

In LTE \cite{3gpp.36.300}, OFDMA technology is used in the downlink, where the channel bandwidth is divided into $15$kHz OFDM subcarriers. The channel bandwidth is intended to be scalable in LTE, and the wider the channel bandwidth, the higher number of OFDM subcarriers are available at the BS. The aggregation of $12$ adjacent OFDM subcarriers and $6$ or $7$ OFDM symbols is referred to as a resource block (RB). Each RB spans over $180$kHz on the frequency axis and $0.5$ms on the time axis. An RB is the smallest resource structure that is given to a user for possible transmission. The number of RBs available at a given BS depends on the channel bandwidth and scheduling interval duration at that BS. The number of RBs allocated to each user on the other hand, depends on the quality of service the user requires. For instance, if a user requires a high data rate, the number of RBs allocated to that user is higher than that of a user requiring less data rate. As an another example, if the QoS is defined on the outage probability, higher number of RBs increases the throughput of a user linearly and decreases the likelihood of outage. In order to model today's cellular technology more realistically, we work with this notion of RBs, and assume that each BS $j \in \mathcal{B}$ has access to $N_j$ RBs to distribute among its associated users. 
\subsection{The channel model, instantaneous rate, long term rate and outage probability}
We denote the positive channel power gain between user $i$ and BSs $j$ by $H_{ij}$, i.e., the received power at user $i$ from BS $j$ is $H_{ij}P_j$. Furthermore, $H_{ij}$ embodies the effects of path loss, log normal shadowing and antenna gains as large scale fading component (denoted by $G_{ij}$), and multi-path Rayleigh fading as small scale fading component (denoted by $F_{ij}$). By adopting these notations we have 
\begin{align}
H_{ij}=G_{ij}F_{ij}, \quad \ \forall (i,j) \in \mathcal{U} \times \mathcal{B}, 
\end{align}
where $\cdot \times \cdot$ denotes the Cartesian product.
The large scale fading component $G_{ij}$ is assumed to be constant during one association period, while the small scale fading component $F_{ij}$ fluctuates fast enough so that a mobile user can average it out in its channel measurements. $F_{ij}$s are modelled by statistically independent exponentially distributed random variables with unit variance. They are exponentially distributed since in Rayleigh fading, the envelope of the signal is assumed to follow a Rayleigh distribution, and in turn the channel power is exponentially distributed \cite{Rappaport}. $F_{ij}$s are statistically independent random variables since they model geographically separated wireless channels which show independent multi-path fading behaviours. Based on these assumptions, $H_{ij}$s are statistically independent exponentially distributed random variables with parameter $\lambda_{ij}$, where
\begin{align}
\lambda_{ij}=\frac{1}{\text{E}[H_{ij}]}=\frac{1}{G_{ij}},
\end{align}
where $\text{E}[\cdot]$ denotes the expected value. As it is mentioned before, user $i$ can measure $G_{ij}$ (and equivalently $\lambda_{ij}$) for all the BSs $j \in \mathcal{B}$.

In such a setting, the instantaneous SINR seen by user $i \in \mathcal{U}$ from BS $j \in \mathcal{B}$ is
\begin{align}
\label{SINR}
%c_{ij}=\log ( 1+\frac{P_j |H_{ij}|^2}{ \sum_{\hat{j} \in \mathcal{B}\setminus j} \big(P_\hat{j} |H_{i \hat{j}}|^2 \big) + N_0 } )
\SINR_{ij}= \frac{P_j H_{ij}}{ \sum_{k \in \mathcal{B} \setminus j }  P_k H_{ik} + B N_0 },
\end{align}
and the long term SINR that is measured by user $i \in \mathcal{U}$ from BS $j \in \mathcal{B}$ is
\begin{align}
\label{long term SINR}
\overline{\SINR}_{ij}= \frac{P_j G_{ij}}{ \sum_{k \in \mathcal{B} \setminus j }  P_k G_{ik} + B N_0 }.
\end{align}
In equations (\ref{SINR}) and (\ref{long term SINR}), the constant $B$ denotes the bandwidth over which an RB is realized, $N_0$ denotes the thermal noise spectral power, and $\mathcal{B} \setminus j$ is the set of all BSs except BS $j$. 

Accordingly, the instantaneous and long term spectral efficiency at user $i$, if it is served by BS $j$, denoted by $c_{ij}$ and $\bar{c}_{ij}$ respectively, can be written as
\begin{align}
\label{capacity}
%c_{ij}=\log ( 1+\frac{P_j |H_{ij}|^2}{ \sum_{\hat{j} \in \mathcal{B}\setminus j} \big(P_\hat{j} |H_{i \hat{j}}|^2 \big) + N_0 } )
c_{ij}=\log_2 ( 1+\SINR_{ij} ), \\
\label{long term capacity}
\bar{c}_{ij}=\log_2 ( 1+\overline{\SINR}_{ij} ).
\end{align}
Without loss of generality, $c_{ij}$ and $\bar{c}_{ij}$ can be regarded as achievable rate and long term achievable rate on an RB. For example, if $c_{ij}$ is multiplied by the RB bandwidth and time duration and divided by the scheduling interval, it will be the achievable rate on one RB. 

Given that $n_{ij}$ RBs are given to user $i$ by BS $j$, the instantaneous and long term data rates seen by user $i$ are
\begin{align}
\label{rate}
r_{ij}=n_{ij}c_{ij},  \\
\label{long term rate}
\bar{r}_{ij}=n_{ij}\bar{c}_{ij}.
\end{align}

We define the outage event of a single user $i$ served by BS $j$ to be the event where the instantaneous rate seen by user $i$ drops below a certain threshold $\gamma_i$. We denote the probability of this event by $P_{ij}^{\text{out}}$ and formally define it as
\begin{align}
\label{single_outage1}
&P_{ij}^{\text{out}}=\Pr\{r_{ij} < \gamma_i \}  \\ \nonumber
&=\Pr \bigg\{ n_{ij}\log_2 \bigg(1+\frac{P_j|H_{ij}|^2}{\sum_{k \in \mathcal{B} \setminus j }  P_{k} H_{ik} + B N_0 }\bigg) \leq \gamma_i \bigg\},
\end{align}
%\begin{align}
%\label{single_outage1}
%P_{ij}^{\text{out}}=\Pr\{r_{ij} < \gamma_i \}  =\Pr \bigg\{ n_{ij}\log_2 \bigg(1+\frac{P_j|H_{ij}|^2}{\sum_{k \in \mathcal{B} \setminus j }  P_{k} H_{ik} + B N_0 }\bigg) \leq \gamma_i \bigg\},
%\end{align}
where $\Pr\{\cdot\}$ denotes the probability of the input argument. This probability of outage is derived in \cite{Boyd_outage,Yao_outage} (an easy to read proof is available in \cite{Boyd_outage}), which is
\begin{multline}
\label{single_outage2}
P_{ij}^{\text{out}}= \\ 
1-\bigg( e^{\frac{-\lambda_{ij}BN_0}{P_j} \SINR^{\text{th}}_{ij}} \bigg)\prod_{k \in \mathcal{B} \setminus j}  \Bigg(  \frac{\frac{\lambda_{ik}}{P_{k}}}{\frac{\lambda_{ij}}{P_j}\SINR_{ij}^{\text{th}}+\frac{\lambda_{ik}}{P_k} }  \Bigg),
\end{multline}
%\begin{align}
%\label{single_outage2}
%P_{ij}^{\text{out}}= 
%1-\bigg( e^{\frac{-\lambda_{ij}BN_0}{P_j} \SINR^{\text{th}}_{ij}} \bigg)\prod_{k \in \mathcal{B} \setminus j}  \Bigg(  \frac{\frac{\lambda_{ik}}{P_{k}}}{\frac{\lambda_{ij}}{P_j}\SINR_{ij}^{\text{th}}+\frac{\lambda_{ik}}{P_k} }  \Bigg),
%\end{align}
where
\begin{align}
\SINR^{\text{th}}_{ij}=2^{\frac{\gamma_i}{n_{ij}}}-1.
\end{align}
Note that $P_{ij}^{\text{out}}$  is measurable by user $i$ since users can measure $\lambda_{ij}$ for all the BSs $j \in \mathcal{B}$ . Moreover, 
It can easily be verified that $P_{ij}^{\text{out}}$ is a strictly decreasing function of $n_{ij}$, i.e., more RBs improves the outage behaviour.

\subsection{Rate and outage QoS constraints}
\label{subsect:QoS}
In this paper, we define two QoS constraints, namely long term rate QoS and outage QoS constraints. We refer to the long term QoS constraint simply as rate QoS constraint hereafter. In the case of rate QoS constraint, each user intends to keep its \emph{long term rate} above its requested rate threshold. In the beginning phase of cell association, each user $i$ requests a certain rate QoS class in terms of minimum required long term rate $\gamma_i$. Therefore, if the user $i$ is associated with BS $j$, it is the duty of the BS to satisfy the following rate QoS constraint
\begin{align}
\label{rQoS1}
\bar{r}_{ij} \geq \gamma_i.
\end{align}
By substituting (\ref{long term rate}) in the above equation, we have
\begin{align}
\label{rQoS2}
n_{ij} \geq \frac{\gamma_i}{\bar{c}_{ij}}.
\end{align}
We indicate the smallest integer greater than the right hand side of the above equation by $\bar{n}_{ij}^R$ as follows
\begin{align}
\label{rQoS3}
\bar{n}_{ij}^R=\lceil  \frac{\gamma_i}{\bar{c}_{ij}}  \rceil,
\end{align}
where $\lceil \cdot \rceil$ represents the ceiling function. Inequalities (\ref{rQoS1}) and (\ref{rQoS2}) and equality (\ref{rQoS3}) indicate that if user $i$ requires rate QoS class of minimum rate $\gamma_i$, BS $j$ will be obliged to allocate at least $\bar{n}_{ij}^R$ RBs to that user, i.e.,
\begin{align}
\label{rQoS4}
n_{ij} \geq \bar{n}_{ij}^R.
\end{align}

In the case of outage QoS constraint, each user intends to keep its \emph{instantaneous rate} above its requested rate threshold with a certain probability. In the beginning of cell association phase, each user requests a certain outage QoS class. An outage QoS class is defined in terms of user's rate threshold $\gamma_i$, and probability of user's rate dropping below that threshold $T_i$. Therefore, if user $i$ is associated with BS $j$, it is the duty of BS $j$ to satisfy the following constraint for that user:
\begin{align}
\label{oQoS1}
\big( P_{ij}^{\text{out}}=\text{Pr} \{r_{ij} \leq \gamma_i \} \big) \leq T_i.
\end{align}
The probability of outage is given in (\ref{single_outage2}) as a function of $n_{ij}$. Since this probability is a strictly decreasing function of $n_{ij}$, a lower bound on $n_{ij}$ exists above which the outage QoS constraint is satisfied. Since $n_{ij}$s can take only positive integer values, this lower bound can easily be found numerically by setting $n_{ij}=1$ in (\ref{single_outage2}) and incrementing it until the constraint is satisfied. We indicate the smallest integer for which (\ref{oQoS1}) is satisfied by $\bar{n}_{ij}^O$. Therefore, if user $i$ requires outage QoS class of rate threshold $\gamma_i$ and probability of outage $T_i$, BS $j$ will be obliged to allocate at least $\bar{n}_{ij}^O$ RBs to that user, i.e.,
\begin{align}
\label{oQoS2}
n_{ij} \geq \bar{n}_{ij}^O.
\end{align}
%%%%%%%%%%%%%%%%%%%%%%%%%%%%%%%%%%%%%%%%%%%%%%%%%%%%%%%%%%%%%%%%%%%%%%%%%%%%%%%%%%%
%%%%%%%%%%%%%%%%%%%%%%%%%%%%%%%%%%%%%%%%%%%%%%%%%%%%%%%%%%%%%%%%%%%%%%%%%%%%%%%%%%%
%%%%%%%%%%%%%%%%%%%%%%%%%%%%%%%%%%%%%%%%%%%%%%%%%%%%%%%%%%%%%%%%%%%%%%%%%%%%%%%%%%%
\section{Problem formulation}
Two optimization problems are considered in this paper: sum utility of long term rate maximization with rate QoS constraints (referred to as $\mathbf{P1}$), and global outage probability minimization with outage QoS constraints (referred to as $\mathbf{P2}$), both through cell association and RB allocation. We formulate these two problems in the rest of this section. Before proceeding further, we define binary association indices $x_{ij}  \in \{0,1\}, \; \forall (i,j) \in \mathcal{U} \times \mathcal{B}$, where $x_{ij}=1$ indicates that user $i$ is associated with BS $j$, and $x_{ij}=0$ indicates the opposite.

\subsection{Sum utility of long term rate maximization with rate QoS constraints}
In this problem, the objective is to maximize a function of long term rate while satisfying the rate QoS constraints. Since we are working with the notion of long term rate, this framework is well-suited for environments with users with low mobility so that the channels remain unchanged in one resource allocation period. We select the sum utility of users' long term rate to be our objective function. Utility of rate can be regarded as a measure of user's satisfaction with the rate it gets. A utility function, in general, is a strictly increasing and concave function. For instance, logarithm function is a suitable candidate. However, in order to preserve generality, the notion of $U(\cdot)$ is used as a general strictly increasing and concave utility function. The sum utility of long term rate maximization problem with rate QoS provision ($\mathbf{P1}$) is
\begin{align}
 \mathbf{P1}:\;\underset{\mathbf{x}, \mathbf{n}}{\text{maximize}} \quad
 & \sum_{i \in \mathcal{U}} \sum_{i \in \mathcal{B}} x_{ij} U(\bar{r}_{ij})  \label{Objective_Function} \\
 \text{subject to} \hspace{7pt}
&\text{(RC)}: \; \sum_{i \in \mathcal{U}} x_{ij}n_{ij} \leq N_j, \; \forall j \in \mathcal{B} \nonumber \\
&\text{(AC)}: \; \sum_{j \in \mathcal{B}} x_{ij} \leq 1, \; \forall i \in \mathcal{U}  \nonumber \\
&\sum_{j \in \mathcal{B}} x_{ij} \bar{r}_{ij} \geq \gamma_i, \;\forall i \in \mathcal{U} \label{rQoS constraint} \\
&x_{ij}  \in \{0,1\}, \; \forall (i,j) \in \mathcal{U} \times \mathcal{B} \label{x binary constraint}\\
&n_{ij}  \in \{0, 1, \dots, N_j\} , \; \forall (i,j) \in \mathcal{U} \times \mathcal{B} \label{n integer constraint}.
\end{align}
In the above optimization problem, $\mathbf{x}$ and $\mathbf{n}$ are matrices containing $x_{ij}$ and $n_{ij}$ elements. Furthermore, the first constraint is referred to as the resource constraint (RC). This constraint ensures that the number of RBs given to the associated users does not exceed the resource budget of that BS. The second constraint is referred to as association constraint (AC). This constraint guarantees that each user is connected to at most one BS. The third constraint (\ref{rQoS constraint}) is the rate QoS constraint which is derived based on inequality (\ref{rQoS1}). In the end, constraints (\ref{x binary constraint}) and (\ref{n integer constraint}) indicate that the association indices are binary variables, and $n_{ij}$s can take integer values between zero and the maximum number of RBs at BS $j$.

\subsection{Global outage probability minimization with outage QoS constraints}
The motivation behind formulating this problem is to take into account the stochastic behaviour of wireless channels without adding signalling overhead to the system. Guaranteeing a constant instantaneous rate to users in wireless environments that suffer from fast fading is not achievable. However, it is possible to guarantee a certain rate with a certain probability, which suggests formulating the problem in the context of outage probability. In the context of outage probability, the eventual goal is to associate the users with BSs and RBs such that the global outage probability is minimized.  In order to achieve this goal, the first step is to define the global outage probability in some sense and evaluate it as a function of system model parameters. We define the global outage event as the event where one or more users experience outage, i.e., if at least one user experiences an instantaneous data rate below its requested threshold, a global outage will be declared. The outage probability of a single link is given in (\ref{single_outage2}). Considering that the outage events for different users are statistically independent, it can be argued that the probability of no users experiencing outage is $\prod_{i \in \mathcal{U}} \prod_{j \in \mathcal{B}} \big(1-P_{ij}^{\text{out}}\big)^{x_{ij}}.$ 
Therefore, the global outage probability indicated by $\widehat{P_{\text{out}}}$ is
\begin{align}
\label{global_outage}
\widehat{P_{\text{out}}}=1-\prod_{i \in \mathcal{U}} \prod_{j \in \mathcal{B}} \big(1-P_{ij}^{\text{out}}\big)^{x_{ij}}.
\end{align}
Now, we define the global outage probability minimization problem with outage QoS provision ($\mathbf{P2}$) as
\begin{align}
 \mathbf{P2}: \; \underset{\mathbf{x}, \mathbf{n}}{\text{minimize}} \hspace{7pt}
 &1-\prod_{i \in \mathcal{U}} \prod_{j \in \mathcal{B}} \big(1-P_{ij}^{\text{out}}\big)^{x_{ij}}  \label{Objective_Function} \\
 \text{subject to} \hspace{7pt}
 &\text{(RC), (AC), (\ref{x binary constraint}), (\ref{n integer constraint})}, \nonumber \\
&\prod_{j \in \mathcal{B}} \text{Pr} \{n_{ij}c_{ij}  \leq \gamma_i \}^{x_{ij}} \leq T_i, \;\forall i \in \mathcal{U} \label{oQoS constraint}.
\end{align}
Comparing the constraints in $\mathbf{P2}$ and $\mathbf{P1}$, the only different constraint is the QoS constraint; in $\mathbf{P2}$ the outage QoS constraint has replaced the rate QoS constraint in $\mathbf{P1}$ for each user. Constraint (\ref{oQoS constraint}) is derived based on the inequality (\ref{oQoS1}).

%%%%%%%%%%%%%%%%%%%%%%%%%%%%%%%%%%%%%%%%%%%%%%%%%%%%%%%%%%%%%%%%%%%%%%%%%%%%%%%%%%%
%%%%%%%%%%%%%%%%%%%%%%%%%%%%%%%%%%%%%%%%%%%%%%%%%%%%%%%%%%%%%%%%%%%%%%%%%%%%%%%%%%%
%%%%%%%%%%%%%%%%%%%%%%%%%%%%%%%%%%%%%%%%%%%%%%%%%%%%%%%%%%%%%%%%%%%%%%%%%%%%%%%%%%%
\section{Cell association phase}
\label{sec:Cell association}
Optimization problems $\mathbf{P1}$ and $\mathbf{P2}$ are combinatorial problems in $\mathbf{x}$ and $\mathbf{n}$ which are involved to solve. In order to make the problem tractable, we solve them in two steps. First, we fix $n_{ij}$s and find association indices $x_{ij}$s. This step is equivalent to solving the association problem. In the next step, given the association indices, we solve the optimization problem with respect to $n_{ij}$s. In this section, we address the association problem, while optimizing with respect to $n_{ij}$s is addressed in the next section. It should be noted that $n_{ij}$s can not be fixed at arbitrary values since the QoS constraints need to be satisfied in the cell association phase. Therefore, we replace the rate QoS constraints in $\mathbf{P1}$ by $n_{ij}=\bar{n}_{ij}^{R}$, and outage QoS constraints in $\mathbf{P2}$ by $n_{ij}=\bar{n}_{ij}^{O}$, for all $(i,j) \in \mathcal{U} \times \mathcal{B}$. According to inequalities (\ref{rQoS4}) and (\ref{oQoS2}), $\bar{n}_{ij}^{R}$ and $\bar{n}_{ij}^{O}$ are the minimum number of RBs required by user $i$ from BS $j$ to satisfy the rate QoS and outage QoS constraints, respectively.

From another perspective, if we replaced $n_{ij}$s by constant $1$ in the (RC), it could be shown that both problems would become a two dimensional assignment problem with respect to $x_{ij}$s, and algorithms such as Hungarian method would solve them efficiently \cite{Hungarian} in a centralized fashion. However, optimizing over $x_{ij}$s is an NP-hard problem because of the (RC) constraint \cite{pentico2007assignment}. In order to change the combinatorial nature of the problem into a continuous one, and hopefully a convex program, we relax the constraints (\ref{x binary constraint}) in both optimization problems, i.e., we replace the constraints (\ref{x binary constraint}) by $0 \leq x_{ij} \leq 1$ for all $(i,j) \in \mathcal{U} \times \mathcal{B}$.

Next, we show that by fixing $n_{ij}$s, problems $\mathbf{P1}$ and $\mathbf{P2}$ become equivalent optimization problems. After the aforementioned modifications, the problem $\mathbf{P1}$ transforms into problem $\mathbf{P1}_\mathbf{x}$ as follows
\begin{align}
\mathbf{P1}_\mathbf{x}: \; \underset{\mathbf{x}}{\text{maximize}} \quad
 &\sum_{i \in \mathcal{U}} \sum_{j \in \mathcal{B}} x_{ij}a_{ij}^{R}\\
 \text{subject to} \hspace{7pt}
& \text{(RC), (AC),}   \nonumber \\
&n_{ij} = \bar{n}_{ij}^{R}, \; \forall (i,j) \in \mathcal{U} \times \mathcal{B}, \\
&0 \leq x_{ij} \leq 1 , \; \forall (i,j) \in \mathcal{U} \times \mathcal{B},
\end{align}
where $$a_{ij}^{R}=U(\bar{r}_{ij})=U(\bar{n}_{ij}^R \bar{c}_{ij}).$$ 
$a_{ij}^{R}$ is treated as a constant since $\bar{n}_{ij}^R$ is a constant.

It seems that the objective function in $\mathbf{P2}$ has a different structure compared to the objective function in $\mathbf{P1}$. However, applying the following changes unifies these two objective functions. Inspecting the objective function (\ref{Objective_Function}), it can be seen that the first term is a constant and can be removed. Moreover, removing the negative sign changes the minimization to a maximization problem. Finally, taking the natural logarithm of the objective function does not change the optimum argument and transforms multiplication to addition. After these modifications, the problem $\mathbf{P2}$ transforms into problem $\mathbf{P2}_\mathbf{x}$ as follows
\begin{align}
\mathbf{P2}_\mathbf{x}: \; \underset{\mathbf{x}}{\text{maximize}} \quad
 &\sum_{i \in \mathcal{U}} \sum_{j \in \mathcal{B}} x_{ij}a_{ij}^{O}\\
 \text{subject to} \hspace{7pt}
& \text{(RC), (AC),}   \nonumber \\
&n_{ij} = \bar{n}_{ij}^{O}, \; \forall (i,j) \in \mathcal{U} \times \mathcal{B}, \\
&0 \leq x_{ij} \leq 1 , \; \forall (i,j) \in \mathcal{U} \times \mathcal{B},
\end{align}
where $$a_{ij}^{O}=\log \big(1-P_{ij}^{\text{out}} \big).$$ 
$a_{ij}^{O}$ is also treated as a constant since $P_{ij}^{\text{out}}$ is a function of $n_{ij}$ which is set to $\bar{n}_{ij}^O$.

By comparing $\mathbf{P1}_\mathbf{x}$ and $\mathbf{P2}_\mathbf{x}$, it is trivial that these two optimization problems have the same structure. Therefore, we remove the superscripts $R$ and $O$ from $a_{ij}^R$, $a_{ij}^O$, $\bar{n}_{ij}^R$, and $\bar{n}_{ij}^O$ and replace them by $a_{ij}$ and $\bar{n}_{ij}$ to have the unified optimization problem $\mathbf{P}_\mathbf{x}$ as follows
\begin{align}
\mathbf{P}_\mathbf{x}: \; \underset{\mathbf{x}}{\text{maximize}} \quad
 &\sum_{i \in \mathcal{U}} \sum_{j \in \mathcal{B}} x_{ij}a_{ij}\\
 \text{subject to} \hspace{7pt}
&   \sum_{i \in \mathcal{U}} x_{ij}\bar{n}_{ij} \leq N_j, \; \forall j \in \mathcal{B}, \label{RC2}\\
&   \sum_{j \in \mathcal{B}} x_{ij} \leq 1, \; \forall i \in \mathcal{U}, \\
&0 \leq x_{ij} \leq 1 , \; \forall (i,j) \in \mathcal{U} \times \mathcal{B}.
\end{align}
The objective function of $\mathbf{P}_\mathbf{x}$ is a linear function in $x_{ij}$s, and all the constraints are linear and affine in $x_{ij}$s. Therefore, $\mathbf{P}_\mathbf{x}$ is a convex optimization problem with respect to $x_{ij}$s \cite{boyd2004convex}.

\subsection{Cell association solution}
In order to devise a distributed solution to the $\mathbf{P}_\mathbf{x}$, we use a similar to the approach suggested in \cite{Andrews-association}, that is employing the Lagrange dual decomposition method \cite{low1999optimization}. Note that the strong duality property holds for  $\mathbf{P}_\mathbf{x}$, that is the optimum value of $\mathbf{P}_\mathbf{x}$ is equal to the optimum value of its Lagrange dual function. According to Slater's theorem, strong duality holds for a convex optimization problem if Slater's condition holds for the constraints of that problem. Moreover, if the problem is convex and all the equality and inequality constraints are linear and affine, Slater's condition reduces to feasibility condition \cite{boyd2004convex}. As discussed in the previous subsection, $\mathbf{P}_\mathbf{x}$ is a convex optimization problem with linear and affine constraints. Therefore, if a set of $x_{ij}$s exists for which $\mathbf{P}_\mathbf{x}$ is feasible, then the strong duality holds. For instance, $x_{ij}=0$, for all $(i,j) \in \mathcal{U} \times \mathcal{B}$ is always a feasible point in  $\mathbf{P}_\mathbf{x}$, thus, strong duality always holds for  $\mathbf{P}_\mathbf{x}$.

We define the Lagrangian of  $\mathbf{P}_\mathbf{x}$ by taking the resource constraint (\ref{RC2}) inside the objective function and indicate it by $\mathcal{L}(\mathbf{x}, \mathcal{\mu})$. The Lagrangian of $\mathbf{P}_\mathbf{x}$ is
\begin{align}
\label{Lagrangian}
\mathcal{L}(\mathbf{x}, \mathbf{\mu})= \sum_{i \in \mathcal{U}} \sum_{j \in \mathcal{B}} x_{ij}a_{ij} - \sum_{j \in \mathcal{B}} \mu_j \big( \sum_{i \in \mathcal{U}} x_{ij}\bar{n}_{ij} - N_j \big),
\end{align}
where $\mu_j$s are Lagrange multipliers associated with resource constraints at BSs. Then, the Lagrange dual function represented by $g(\mathbf{\mu})$ is
\begin{align}
g(\mathbf{\mu})=\underset{\mathbf{x}}{\sup} \quad & \sum_{i \in \mathcal{U}} \sum_{j \in \mathcal{B}} x_{ij}(a_{ij} - \mu_j \bar{n}_{ij})+\sum_{j \in \mathcal{B}}\mu_j N_j\\
\text{subject to} \hspace{7pt}
&0 \leq x_{ij} \leq 1 , \; \forall (i,j) \in \mathcal{U} \times \mathcal{B}, \\
&\sum_{j \in \mathcal{B}} x_{ij} \leq 1, \; \forall i \in \mathcal{U}.
\end{align}
The strong duality holds, therefore, we can first maximize over $\mathbf{x}$ and then minimize over $\mathbf{\mu}$. In order to find $g(\mu)$ for fixed $\mu_j$s, we rewrite the Lagrange dual function as
\begin{align}
g(\mathbf{\mu})=\sum_{i \in \mathcal{U}} g_i(\mathbf{\mu})+ \sum_{j \in \mathcal{B}} \mu_jN_j,
\end{align}
where $g_i(\mathbf{\mu})$ is defined for all $i \in \mathcal{U}$ as
\begin{align}
\label{user_problem}
g_i(\mathbf{\mu})=\underset{x_{ij}, j \in \mathcal{B}}{\sup} \quad  & \sum_{j \in \mathcal{B}} x_{ij}(a_{ij} - \mu_j \bar{n}_{ij})\\
\text{subject to} \hspace{7pt}
&0 \leq x_{ij} \leq 1 , \; j \in  \mathcal{B} \nonumber \\
&\sum_{j \in \mathcal{B}} x_{ij} \leq 1. \nonumber
\end{align}
It can be seen that for fixed $\mu_j$s, $g(\mathbf{\mu})$ is separable with respect to users $i$ as in (\ref{user_problem}). Therefore, each user $i$ needs to solve the optimization problem  (\ref{user_problem}). For a given user $i$, the objective function in (\ref{user_problem}) is a weighted average of $(a_{ij} - \mu_j \bar{n}_{ij})$, where the weights are between $0$ and $1$ and they sum up to unity. Therefore, (\ref{user_problem})'s unique solution is yielded by keeping the maximum argument of  $(a_{ij} - \mu_j \bar{n}_{ij})$ over $j$s and diminishing the contribution of other elements. We call the term $(a_{ij} - \mu_j \bar{n}_{ij})$ the qualification index of BS $j$ from user's $i$ point of view and indicate it by $\text{QI}_{ij}$ as follows
\begin{align}
\label{qualification index}
\text{QI}_{ij}=a_{ij} - \mu_j \bar{n}_{ij}.
\end{align}
Accordingly, (\ref{user_problem})'s unique solution for each user $i$ is
\begin{align}
\label{association update}
x_{ij}=\begin{cases}
      1 & \text{if } j=j^{*} \\
      0 & \text{if } j \neq j^{*}
    \end{cases}, \; \forall i \in \mathcal{U},
\end{align}
where
\begin{align}
\label{user update}
j^{*}=\underset{j\in \mathcal{B}}{\text{argmax}} (\text{QI}_{ij}), \; \forall i \in \mathcal{U}.
\end{align}
After finding $x_{ij}$s for fixed $\mu_j$s, we update the vector $\mathbf{\mu}$ by using gradient descent method \cite{boyd2004convex}. The partial derivative of the Lagrange dual function with respect to $\mu_j$ is
\begin{align}
\frac{\partial \mathcal{L}(\mathbf{x}, \mathbf{\mu})}{\partial \mu_j}=N_j-\sum_{i \in \mathcal{U}} x_{ij}n_{ij}, \; \forall j \in \mathcal{B}.
\end{align}
Therefore, the updating rule for $\mu$ is
%\begin{multline}
%\label{mu_update}
%\mu_j(t+1)=\bigg[ \mu_j(t)-\beta(t)\big(N_j-\sum_{i \in \mathcal{U}} x_{ij}n_{ij}\big) \bigg]^+, \\
%\forall j \in \mathcal{B},
%\end{multline}
\begin{align}
\label{mu_update}
\mu_j(t+1)=\bigg[ \mu_j(t)-\beta(t)\big(N_j-\sum_{i \in \mathcal{U}} x_{ij}n_{ij}\big) \bigg]^+, 
\forall j \in \mathcal{B},
\end{align}
where the operator $[\cdot]^+$ indicates the maximum of the argument of the operator and $0$. We applied the operator $[\cdot]^+$ on $\mu_j$s because the Lagrange multipliers are non-negative parameters \cite{boyd2004convex}. Furthermore, the $\beta(t)$ is some step size that satisfies the following two conditions
\begin{align}
\lim_{t \rightarrow \infty} \beta(t) = 0 \quad, \text{ and} \quad  \sum_{t=1}^{\infty}\beta(t)=\infty.
\end{align}
 According to proposition 6.3.4 in \cite{Bertsekas:2009}, if the step size $\beta(t)$ satisfies the above conditions, the convergence of the gradient descent method will be guaranteed assuming that the association indices are continuous variables of the form $0 \leq x_{ij} \leq 1$. These conditions do not guarantee the convergence for the binary association indices. However, we have used $\beta(t)=0.5/t$ in our simulations, and in all cases the distributed algorithm converged in less than $25$ iterations; most of the times the convergence was reached in less than $10$ iterations. 
 
We iteratively update the association variables according to (\ref{association update}), and the Lagrange multipliers according to (\ref{mu_update}), until the convergence is reached. Note that updating rule for the association variables automatically generates binary values. Consequently, no further approximations is required.

\subsection{Feasibility problem and admission control}
In the cell association procedure described in the previous subsection, the users find their desired BS through (\ref{user update}). The users are not aware of how many other users are associating themselves with the desire BS and how much resources that BS has access to. As a result, many users may associate themselves with a BS and exhaust its resources, leading to violating the resource constraint (\ref{RC2}). This condition not only affects the convergence of the algorithm, but also takes the solution out of the feasible set. As it is described in \cite{boyd2004convex}, the Lagrange dual function is an upper bound on the original maximization problem only if we are in the feasible set of the problem defined by the constraints. Beyond the feasible set, not only the Lagrange dual function may not be an upper bound on the original problem, but also iterating with gradient descent method may divert the solution from the desired optimal point. Therefore, once an iteration is out of the feasible set, it needs to be projected back to the feasible set, i.e., gradient projection method \cite{Bertsekas:1989} should be used.

In the context of cell association in HetNets, we use a huristic to project the iteration back to the feasible set once one or more BSs receive association requests from too many users. First, we require all the users $i\in \mathcal{U}$ to sort the BSs in their range in descending order based on their qualification index $\text{QI}_{ij}$ given in (\ref{qualification index}), and send this sorted list to the BS they want to connect to (BS $j^{*}$ in (\ref{user update})). Let us say BS $j$ receives too many association requests, i.e., upon acceptance of all those requests the resource constraint at BS $j$ is violated. Then, BS $j$ finds the users who are consuming the highest number of RBs (highest number of $\bar{n}_{ij}$), and removes them until its resource constraint is satisfied. BS $j$ sends those removed users' requests to the second best BSs the users have requested. If the resource constraints are satisfied at all BSs now, the projection is accomplished. Otherwise, this procedure continues until the solution is back to the feasible set. As we have mentioned before, it is assumed that BSs are connected through a high speed back-haul, and the message passings required for projecting the solution back to the feasible set takes place in negligible time period. Moreover, the admission control described here achieves load balancing in the network since it avoids over populating the BSs.

\subsection{Distributed cell association protocol design}
The proposed distributed algorithm of cell association described in previous sections is summarized here:

{\bf{Step 1: Initialization}} All BSs $j \in \mathcal{B}$ initialize their associated Lagrange multiplier $\mu_j$ and broadcast them in the network.

{\bf{Step 2: User request}} In this step, all users $i \in \mathcal{U}$  listen to the pilot signals broadcasted by BSs and measure the SINR from each BS and channel gains between themselves and all the BSs. Based on these measurements and the desired QoS, user $i$ calculates the number of RBs it needs from each BS. Then, user $i$ calculates the $\text{QI}_{ij}$ in equation (\ref{qualification index}) for all the BSs in its range and sorts the BSs in descending order. A request containing this sorted list along with the required number of RBs from the BSs in the list is sent to the BS with the best $\text{QI}_{ij}$.

{\bf{Step3: User admission}} BSs process the requests they have received. If BS $j$ can accommodate all the requests it has received, an admission message is sent to all those users finding BS $j$ to be the best candidate. Otherwise, BS $j$ forwards the requests from users consuming the highest amount of resources to the next best BS the users have requested. This procedure continues until the solution is feasible.

{\bf{Step 4: BS Lagrange multiplier update}} After all the users are accommodated, BSs update their Lagrange multipliers according to (\ref{mu_update}) and broadcast the new multipliers. The algorithm continues by going back to step 2.

This algorithm solves the cell association problem in a distributed fashion for either of the rate maximization with rate QoS constraints problem, or global outage probability minimization with outage QoS constraints problem ($\mathbf{P1}_\mathbf{x}$ or $\mathbf{P2}_\mathbf{x}$).   

%%%%%%%%%%%%%%%%%%%%%%%%%%%%%%%%%%%%%%%%%%%%%%%%%%%%%%%%%%%%%%%%%%%%%%%%%%%%%%%%%%%
%%%%%%%%%%%%%%%%%%%%%%%%%%%%%%%%%%%%%%%%%%%%%%%%%%%%%%%%%%%%%%%%%%%%%%%%%%%%%%%%%%%
%%%%%%%%%%%%%%%%%%%%%%%%%%%%%%%%%%%%%%%%%%%%%%%%%%%%%%%%%%%%%%%%%%%%%%%%%%%%%%%%%%%
\section{RB distribution phase}
After the cell association phase is completed, some of the BSs may have extra RBs not allocated to any users. In this section, in order to allocate the remaining RBs, given the association indices $x_{ij}$s we solve the optimization problems $\mathbf{P1}$ and $\mathbf{P2}$ for $n_{ij}$ with fixed $x_{ij}$s, . Before addressing either of $\mathbf{P1}$ and $\mathbf{P2}$, we define $\mathcal{U}_j$ as the set of users associated with BS $j$
\begin{align}
\mathcal{U}_j= \{ i | x_{ij}=1\}, \forall j \in \mathcal{B}.
\end{align}

\subsection{Sum utility of rate maximization}
Assuming fixed $x_{ij}$s, problem $\mathbf{P1}$ is reduced to the following optimization problem at each BS $j$
\begin{align}
\mathbf{P1}_{\mathbf{n},j}: \; \underset{n_{ij}^{'}, i \in \mathcal{U}_j }{\text{maximize}} \quad
 &\sum_{i \in \mathcal{U}_j}  U( n_{ij}\bar{c}_{ij}) \\
 \text{subject to} \hspace{7pt}
&   \sum_{i \in \mathcal{U}_j} n_{ij} =N_j \label{RC3}, \\
& n_{ij}=\bar{n}_{ij}^R + n_{ij}^{'},
\end{align}
where $\bar{n}_{ij}^R$ is the number of already allocated RBs to user $i$ by BS $j$ satisfying the rate QoS constraint of user $i$, and $n_{ij}^{'}$ is the share of user $i$ from the remaining RBs available at BS $j$. The above problem is a convex problem that can be solved through Karush-Kuhn-Tucker  (KKT) conditions \cite{boyd2004convex}. Using the Lagrange multiplier $\nu$ for the resource constraint (\ref{RC3}), the solution is
\begin{align}
n_{ij}^{'}=\bigg[\frac{1}{\bar{c}_{ij}}(U^{'})^{-1}\big(  \frac{\nu}{\bar{c}_{ij}} \big) -\bar{n}_{ij}^R\bigg]^+, \; \forall i \in \mathcal{U}_j,
\end{align}
where $\nu$ is the unique solution of the following equation
\begin{align}
\sum_{i \in \mathcal{U}_j} \max \bigg\{\frac{1}{\bar{c}_{ij}}(U^{'})^{-1}\big(  \frac{\nu}{\bar{c}_{ij}}, \big),\bar{n}_{ij}^{R}\bigg\} = N_j,
\end{align}
 and $(U')^{-1}(\cdot)$ is the inverse of the derivative of $U(\cdot)$ with respect to $n_{ij}$. The solution of the above equation is unique since $U(\cdot)$ is a concave and strictly increasing function, hence, $(U^{'})^{-1}(\cdot)$ is a strictly increasing (monotonic) function of $\nu$. This equation can be solved efficiently through a numerical search method. In the end, the optimal solution of the $\mathbf{P1}_{\mathbf{n},j}$ is rounded to the closest integer value since the the procedure described here does not necessarily produce integer values for $n_{ij}$s.
 
In our simulations, we chose $U(x)=\log(1+x)$. In this case, $n_{ij}^{'}=\big[{1}/{\nu}-{1}/{\bar{c}_{ij}} - \bar{n}_{ij}^R\big]^+$.

\subsection{Global outage probability minimization}
Assuming fixed $x_{ij}$s, removing the constant 1 in the objective function of $\mathbf{P2}$, flipping the negative sign to positive sign, and taking the logarithm of the remaining term, $\mathbf{P2}$ reduces to the following optimization problem at each BS $j$
\begin{align}
\mathbf{P2}_{\mathbf{n},j}: \; \underset{n_{ij}^{'}, i \in \mathcal{U}_j }{\text{maximize}} \quad
 &\sum_{i \in \mathcal{U}_j}  \log \big(  1-P_{ij}^{\text{out}}\big)  \\
 \text{subject to} \hspace{7pt}
&   \sum_{i \in \mathcal{U}_j} n_{ij} =N_j \label{RC4}, \\
& n_{ij}=\bar{n}_{ij}^O + n_{ij}^{'},
\end{align}
where $\bar{n}_{ij}^O$ is the number of already allocated RBs to user $i$ by BS $j$ satisfying the outage QoS constraint of user $i$, and $n_{ij}^{'}$ is the share of user $i$ from the remaining RBs available at BS $j$. The optimization problem $\mathbf{P2}_{\mathbf{n},j}$ is a non-convex problem, and finding the closed form solution to it is involved. However, it can be shown that the objective function in $\mathbf{P2}_{\mathbf{n},j}$ is strictly increasing in $n_{ij}$s. In other words, increasing each of $n_{ij}$s or a subset of $n_{ij}$s increases the objective function. As this problem is a monotonic combinatorial optimization problem, applying a greedy algorithm is a natural approach to solving it \cite{leiserson2001introduction}. We propose a greedy algorithm where in each iteration one RB is given to the user that benefits the most in terms of the outage probability (the user that has the highest decrement in $P_{ij}^{\text{out}}$ given in (\ref{single_outage2})) until the RBs at BS $j$ are exhausted. At BS $j$, given that there are $K_j=(N_j-\sum_{i \in \mathcal{U}_j}\bar{n}_{ij}^{O})$ RBs left, the algorithm terminates in $K_j$ iterations. 

%However, by assuming high SINR regime and using the one-to-one change of variables of the form ${1}/{y_{ij}}=2^{\frac{\gamma_i}{n_{ij}}}$ in (\ref{single_outage1}), it can be shown that $\mathbf{P2}_{\mathbf{n},j}$ converts to a convex optimization problem if $n_{ij}\geq 1$. Given the high processing power of BSs, this convex optimization problem can be solved through efficient numerical algorithms at the BSs. In the end, the optimal solution of the $\mathbf{P1}_{\mathbf{n},j}$ is rounded to the closest integer value, since the the procedure described here does not necessary produce integer values for $n_{ij}$s.

\section{Numerical results}
In this section, we evaluate the performance of our distributed cell association algorithm and the effects of distributing the remaining RBs through numerical simulations. Three tiers of BSs are considered to exist in the HetNet. The transmitting powers of macro, micro and femto BSs are set to $46, 35$ and $20$ dBm, respectively. The macro BSs' locations are assumed to be fixed and for each macro BS, 5 micro BSs, 10 femto BSs, and 200 users are randomly located in a square area of $1000$ m$\times1000$ m, unless stated otherwise. Regarding the channel model, large scale path loss and small scale Rayleigh multi-paths fading are considered. The path loss between the macro or micro BSs, and the users is modelled as $L(d)=34+40\log_{10}(d)$, and the pass loss between femto BSs and users is $L(d)=37+30\log_{10}(d)$, where $d$ is the distance between users and BSs in meters. The small scale fading is modelled by statistically independent exponentially distributed random variables with unit variance. The noise power at all the receivers is set to $-111.45$ dBm, which corresponds to thermal noise at room temperature and bandwidth of $180$ kHz (Bandwidth of an RB in LTE standard). The mobile users in their SINR and channel gain measurements average out the Rayleigh multi-paths fading and see the effects of large scale path loss, while their instantaneous rate depends on both the large scale and small scale fading. The number of RBs available at macro, micro and femto BSs are $N_{\text{macro}}=200$, $N_{\text{micro}}=100$, and $N_{\text{femto}}=50$. Without loss of generality, the scheduling interval of $1$ second is considered in the simulations.     

\subsection{Rate cumulative distribution functions}
\label{subsect:CDF}
\begin{figure}[tb]
\label{CDF1}
\centering
\includegraphics[width=3.5in]{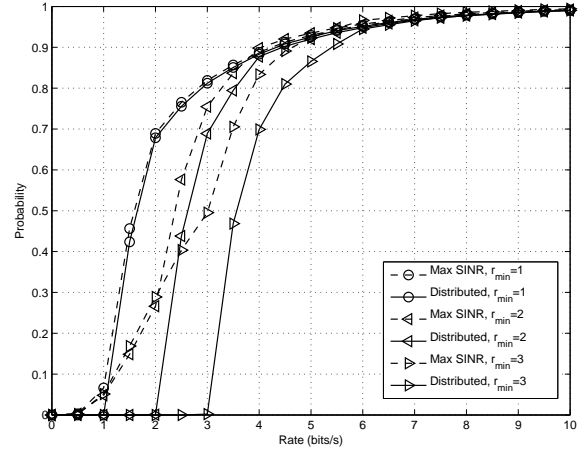}
\caption{The CDFs of users' long term rate for the rate problem in a static setting}
\end{figure}
\begin{figure}[tb]
\label{CDF2}
\centering
\includegraphics[width=3.5in]{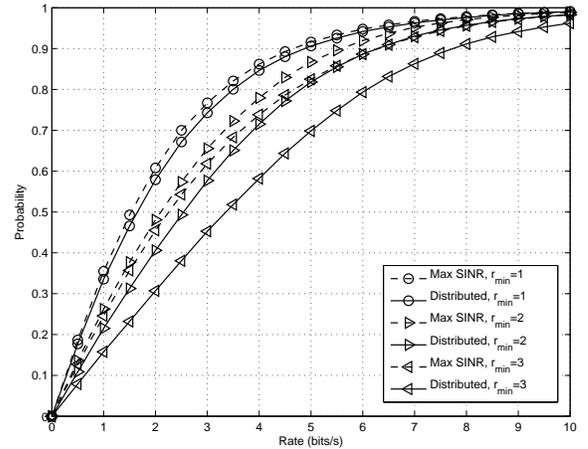}
\caption{The CDFs of users' instantaneous rate for the rate problem in a stochastic setting}
\end{figure}
\begin{figure}[tb]
\label{CDF3}
\centering
\includegraphics[width=3.5in]{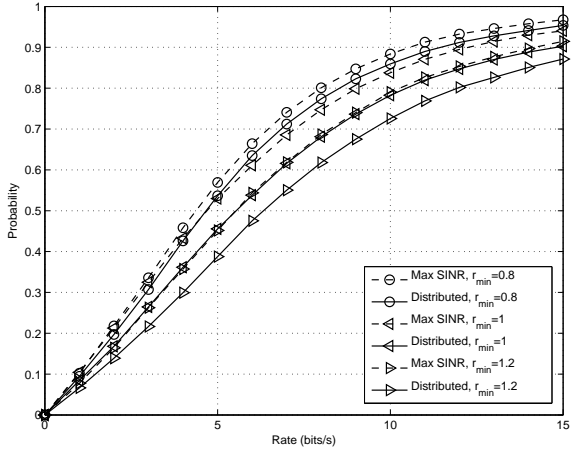}
\caption{The CDFs of users' instantaneous rate for the outage problem and outage probability of $T=10\%$ in a stochastic setting }
\end{figure}
Fig. 1 shows the cumulative distribution function (CDF) of the long term rate for the Max-SINR and the distributed cell association algorithms for the rate problem in a static simulation environment. The simulation environment is static in the sense that the Rayleigh multi-path fading is not considered, and only large scale path loss is taken into account. Fig. 2 shows the CDFs of instantaneous rate for the Max-SINR and the distributed cell association algorithms for the rate problem in a stochastic setting where both large scale and small scale fading are taken into account. In both figures, the results for the rate threshold of $\gamma=1, 2,$ and $3$ bits/s are shown. In the case of Max-SINR  scheme, some of the BSs may get overloaded when users associated with a BS require more RBs than the BS budget; thus, those users are needed to be scheduled in the next scheduling interval. The rate reduction caused by the over-loaded BSs is taken into account in the simulations. As it can be seen in Fig. 1, the long term rate of the users never drops below the rate threshold in the case of the distributed algorithm, while Max-SINR algorithm is not able to satisfy the rate QoS constraints in a static setting. Furthermore, the rate CDFs of the distributed algorithm always lie below the corresponding CDFs resulted by employing the Max-SINR algorithm, especially for the cell edge users (worst $10\%$ users). The rate gain for the cell edge users obtained by using the distributed algorithm over the Max-SINR algorithm increases by increasing the rate threshold. To be more specific, rate gains of $\alpha=2.4, 1.6,$ and $1.1$ are observed for minimum thresholds of $\gamma=3, 2,$ and $1$ bits/s, respectively, in a static setting. Likewise, in a stochastic setting, as it can be seen in Fig. 2, the rate CDFs of the distributed algorithm always lie below the corresponding CDFs obtained by employing the Max-SINR algorithm. In this case, the rate gains for the cell edge users of $\alpha=1.75, 1.3,$ and $1.08$ are observed for minimum thresholds of $\gamma=3, 2,$ and $1$ bits/s. However, the instantaneous rate seen by the users can go below the rate threshold since the users' measurements are based on the average channel gains, while the instantaneous rate is dictated by both the average channel gain and the small scale Rayleigh fading. Since lower rates than the rate threshold are also achievable, the average rates and the rate gain of the cell-edge users drop in the stochastic setting compared to the static setting. 

The CDFs of the instantaneous rate for the Max-SINR and the distributed cell association algorithms for the outage problem is shown in Fig. 3. Only a stochastic setting is considered for the outage problem since outage probability cannot be defined in a static setting. The maximum outage probability is set to $T=10\%$ for all the users. By zooming in this figure, it can be seen that the outage probability for the distributed algorithm and the cases of $\gamma=0.8, 1,$ and $1.2$ bits/s is $T=7.9\%, 8.4\%,$ and $8.1\%$, respectively, which are less than the required outage probability of $T=10\%$; thus, the outage constraints are always satisfied. This is while the Max-SINR algorithm does not necessarily satisfy the outage constraints. Moreover, similar to the rate problem, the CDFs of rate resulted by the distributed algorithm always lie below the CDFs of rate from the Max-SINR algorithm. The rate gain for the cell edge users is $\alpha=1.4, 1.23,$ and $1.1$ for minimum thresholds of $\gamma=1.2, 1,$ and $0.8$ bits/s, respectively. 

\begin{figure}[tb]
\label{Performance}
\centering
\includegraphics[width=3.5in]{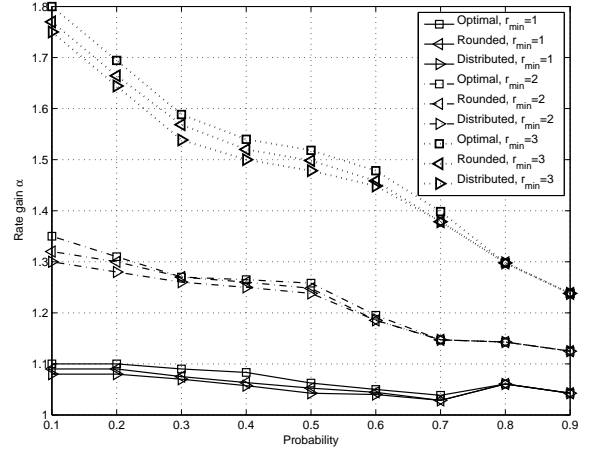}
\caption{The rate gain of the optimal linear program algorithm, the rounding algorithm, and our distributed algorithm over maximum SINR algorithm for the rate problem in a stochastic setting }
\end{figure}
Fig. 4 demonstrates the effectiveness of the distributed cell association algorithm for the rate problem in a stochastic setting. The cell association problem $\mathbf{P}_{\mathbf{x}}$ introduced in section \ref{sec:Cell association} can be solved by three different methods. First one is solving the problem directly as a linear program using the simplex method \cite{linprogBook}. This method does not necessarily produce binary values for the association indices, however, provides an upper bound to all the other methods. We call this method the optimal method. The second method is obtained by rounding the solution of the optimal linear program method to the closest integer value, producing $0$s and $1$s for the association indices. We call this method the rounding method. Finally, we have the distributed algorithm introduced in this paper. In Fig. 4, the rate gain $\alpha_a={{r_a(\text{Probability})} \over {r_{\text{Max-SINR}}(\text{Probability})}}$ is plotted against the probability, where $a \in \{ \text{Optimal}, \text{Rounding}, \text{Distributed}\}$. For instance, at probability $0.2$, $\alpha$ of the optimal algorithm is the ratio of the rate for which {20\%} of the users experience rates below that rate when the problem is solved by the optimal algorithm, over the rate for which {20\%} of the users experience rates below that rate when the problem is solved by the Max-SINR algorithm. It can be seen that the rate gains for the optimal and rounding methods are close, meaning that the solution of the optimal linear program is mostly composed of $0$s and $1$s. Also, the rate gains of the distributed algorithm is close to the rounding algorithm, which proves the effectiveness of our distributed algorithm. We have observed similar trends for the rate problem in a static setting, and outage problem in a stochastic setting. 

%Before concluding this subsection, the convergence of the algorithm is commented. We have used the step size of $\beta(t)=0.5/t$ in updating the Lagrange multipliers $\mu_{j}$ in equation (\ref{mu_update}). In all the cases, the distributed algorithm converged in less than 25 iterations; most of the times the convergence is reached in less than 10 iterations. 
\subsection{The effect of number of femto BSs}
The effects of number of femto BSs per macro BS on the performance of the distributed and Max-SINR algorithms are demonstrated in Figures 5 and 6. In a stochastic setting, Fig. 5 shows the average sum utility of instantaneous rate for the rate problem (problem $\mathbf{P1}_\mathbf{x}$) and rate thresholds of $\gamma=1, 2,$ and $3$ bits/s, while Fig. 6 shows the logarithm of the probability of no users experiencing outage $\log_{10}(1-\widehat{P_{\text{out}}})$ for the outage problem (problem $\mathbf{P2}_\mathbf{x}$) and rate thresholds of $\gamma=0.8, 1,$ and $1.2$ bits/s and outage probability of $T=10\%$. It can be seen that the distributed algorithm outperforms the Max-SINR algorithm in all the cases. It is also observed that the performance of the distributed algorithm slightly worsens by increasing the number of femto BSs, which is attributed to introducing more interference in the network. 

Moreover, for the rate problem in Fig. 5, the performance of the Max-SINR algorithm first improves as the number of femto BSs increases, which is because more resources become available per unit area and the likelihood of having overloaded BSs decreases. As the number of femto BSs surpasses a threshold, the performance of Max-SINR saturates then worsens, similar to the distributed algorithm, since the effect of more interference dominates the more availability of resources. As for the outage problem in Fig. 6, the performance of the Max-SINR algorithm for the cases of $\gamma=1$ and $1.2$ bits/s shows a decreasing trend since the effect of more available RBs never dominates the interference. However, in the case of $\gamma=0.8$ bits/s, first the interference worsens the performance and then the availability of more resources boosts the performance.  
\begin{figure}[tb]
\label{Femto1}
\centering
\includegraphics[width=3.5in]{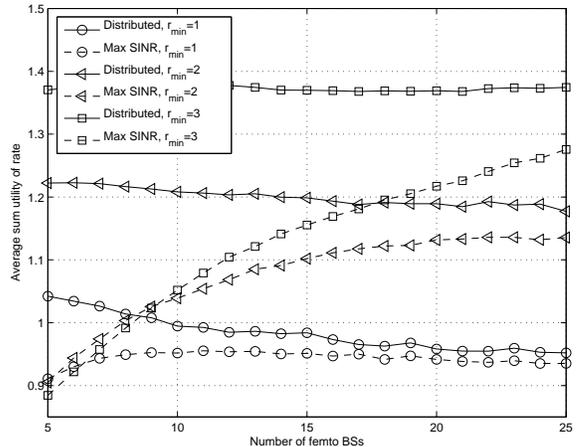}
\caption{The average sum utility of instantaneous rate against number of femto BSs for the rate problem in a stochastic setting}
\end{figure}
\begin{figure}[tb]
\label{Femto2}
\centering
\includegraphics[width=3.5in]{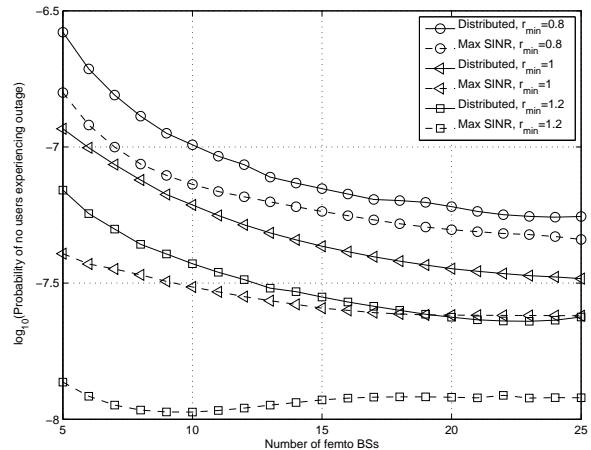}
\caption{The probability of no users being in outage against number of femto BSs for the outage problem and outage probability of $T=10\%$}
\end{figure}
\subsection{The effect of number of users}
\label{subsec:NoOfUsers}
The effects of number of users per macro BS on the performance of the distributed and Max-SINR algorithms are demonstrated in Figures 7 and 8. In a stochastic setting, Fig. 7 shows the average sum utility of instantaneous rate for the rate problem (problem $\mathbf{P1}_\mathbf{x}$) and rate thresholds of $\gamma=1, 2,$ and $3$ bits/s, while Fig. 8 shows the logarithm of the probability of no users experiencing outage $\log_{10}(1-\widehat{P_{\text{out}}})$ for the outage problem (problem $\mathbf{P2}_\mathbf{x}$) and threshold rates of $\gamma=0.8, 1,$ and $1.2$ bits/s and outage probability of $T=10\%$. It can be seen that the distributed algorithm outperforms the Max-SINR algorithm in all the cases. 

For the rate problem, as it can be seen in Fig. 7, the distributed algorithm keeps the average sum utility around a constant value, which is because of the load balancing it achieves. The distributed algorithm provides each user with only enough number of RBs to satisfy the rate constraints. This is why the performance does not vary with the number of users, and increases with increasing the rate threshold. As for the outage problem in Fig. 8, $\log_{10}(1-\widehat{P_{\text{out}}})$ decreases almost linearly. This trend occurs because the distributed algorithm provides users with only enough number of RBs to keep the outage probability of each link slightly below the required outage probability. Besides, more users translates into more multiplicative terms in  $1-\widehat{P_{\text{out}}}=\prod_{i \in \mathcal{U}} \prod_{j \in \mathcal{B}} \big(1-P_{ij}^{\text{out}}\big)^{x_{ij}}$, causing the logarithm of  $1-\widehat{P_{\text{out}}}$ to decrease almost linearly. In an intuitive fashion, when there are more users in the system, the likelihood of at least one user going into outage increases. Therefore, the probability of no users experiencing outage decreases.   

As for the performance of the Max-SINR algorithm, increasing the number of users leads to less availability of resources and decline of the performance in both rate and outage problems. 
\begin{figure}[tb]
\label{User1}
\centering
\includegraphics[width=3.5in]{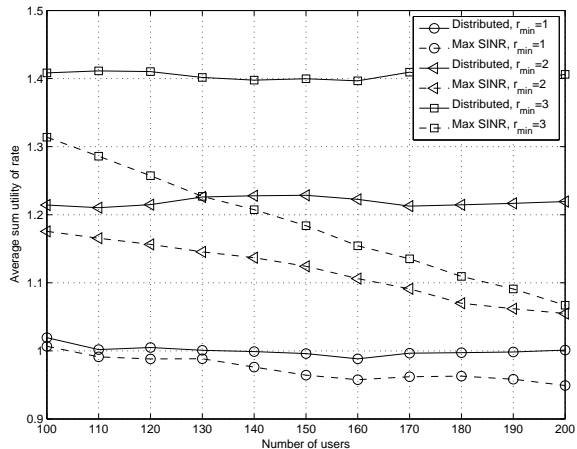}
\caption{The average sum utility of instantaneous rate against number of users for the rate problem in a stochastic setting}
\end{figure}
\begin{figure}[tb]
\label{User2}
\centering
\includegraphics[width=3.5in]{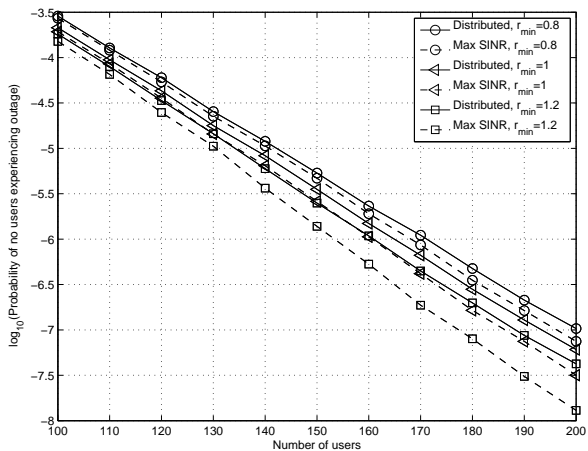}
\caption{The probability of no users being in outage against number of users for the outage problem and outage probability of $T=10\%$}
\end{figure} 

\subsection{The effect of distributing the remaining RBs}
By far, in all our simulations we have considered only the distributed cell association algorithm solving the optimization problem $\mathbf{P}_\mathbf{x}$. The effect of distributing the remaining RBs after the cell association phase for the rate (solving $\mathbf{P1}_{\mathbf{n},j}$ on top of the cell association problem $\mathbf{P1}_\mathbf{x}$) and outage (solving $\mathbf{P2}_{\mathbf{n},j}$ on top of the cell association problem $\mathbf{P2}_\mathbf{x}$) problems is demonstrated in figures 9 and 10, respectively. In Fig. 9, we can see the average sum utility of instantaneous rate for the distributed algorithm solving the cell association problem, and the distributed algorithm with the remaining RBs, against the rate threshold in a stochastic setting. In Fig. 10, we can see the logarithm of the probability of no users experiencing outage for the distributed algorithm , and the distributed algorithm with the remaining RBs, against the rate threshold and link outage probability of $T=10\%$. In both figures, the curves corresponding to 150 and 200 users per macro BS are plotted. The first observation on these two figures is that the distributed algorithm with the remaining RBs significantly outperforms the results obtained through the distributed cell association algorithm only. Secondly, with lower number of users, the performance of the distributed algorithm with the remaining RBs improves. This is because for a given rate threshold, the cell association algorithm first provides users with only enough number of RBs to satisfy the QoS constraints. Therefore, less users require less overall number of RBs to have their QoS constraints satisfied, leaving more RBs unused. More unused RBs trivially translates into a stronger boost in the performance after distributing them among the users. Finally, by increasing the rate threshold, the performance of the distributed cell association only algorithm gets close to the performance of the distributed algorithm with the remanning RBs. This trend is seen since more RBs are required to satisfy QoS constraints with higher rate thresholds, leaving less overall unused RBs in the network. Distributing less unused RBs among users leads to a less improvement over distributed cell association algorithm.  

\begin{figure}[tb]
\label{RB1}
\centering
\includegraphics[width=3.5in]{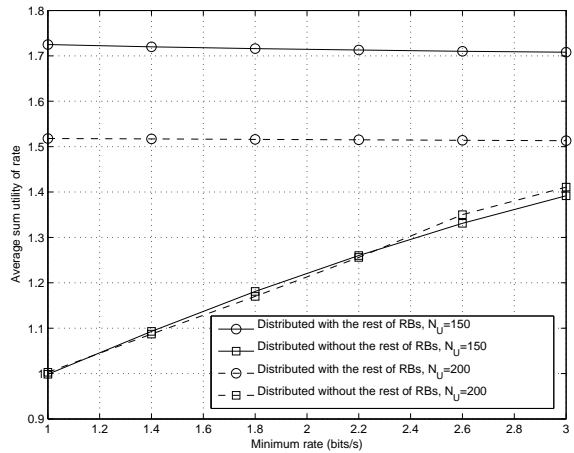}
\caption{The average sum utility of instantaneous rate against minimum rate for the rate problem showing the effect of distributing the remaining RBs in a stochastic setting}
\end{figure}

\begin{figure}[tb]
\label{RB2}
\centering
\includegraphics[width=3.5in]{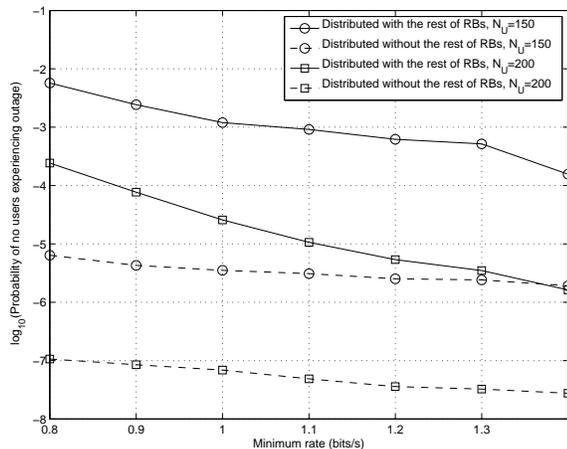}
\caption{The probability of no users being in outage against minimum rate for the outage problem and outage probability of $T=10\%$ showing the effect of distributing the remaining RBs}
\end{figure}

\section{Conclusions}
This paper addressed  the cell association problem in the downlink of a multi-tier HetNet, where BSs have finite number of RBs available to distribute among their associated users. We proposed a QoS-driven distributed cell association algorithm where users receive only enough number of RBs to satisfy their QoS constraints while maximizing the sum utility of long term rate or minimizing the global outage probability. The option of  distributing the remaining RBs is also given to the BSs after the cell association phase. The algorithms derived in this paper are of low complexity, and low levels of message passing is required to render them distributed. Extensive simulation results that are brought in this paper show the superiority of our distributed cell association scheme compared to maximum SINR scheme. For instance, rate gains of up to 2.4x have been observed in the simulations for the cell edge users in our distributed cell association algorithm over maximum SINR scheme. Most importantly, we provided a general framework for jointly associating users to BSs and RBs in LTE systems where frequency reuse factor of $1$ and no interference coordination are assumed. Resource partitioning and incorporating interference avoidance techniques in this framework are among possible future research venues.

\bibliographystyle{IEEEtran}
\bibliography{Boostanimehr_CL2011-2567.R1_bibliography}

\begin{thebibliography}{10}
\providecommand{\url}[1]{#1}
\csname url@samestyle\endcsname
\providecommand{\newblock}{\relax}
\providecommand{\bibinfo}[2]{#2}
\providecommand{\BIBentrySTDinterwordspacing}{\spaceskip=0pt\relax}
\providecommand{\BIBentryALTinterwordstretchfactor}{4}
\providecommand{\BIBentryALTinterwordspacing}{\spaceskip=\fontdimen2\font plus
\BIBentryALTinterwordstretchfactor\fontdimen3\font minus
  \fontdimen4\font\relax}
\providecommand{\BIBforeignlanguage}[2]{{%
\expandafter\ifx\csname l@#1\endcsname\relax
\typeout{** WARNING: IEEEtran.bst: No hyphenation pattern has been}%
\typeout{** loaded for the language `#1'. Using the pattern for}%
\typeout{** the default language instead.}%
\else
\language=\csname l@#1\endcsname
\fi
#2}}
\providecommand{\BIBdecl}{\relax}
\BIBdecl

\bibitem{hasan2011green}
Z.~Hasan, H.~Boostanimehr, and V.~K. Bhargava, ``Green cellular networks: A
  survey, some research issues and challenges,'' \emph{IEEE Commun. Surveys \&
  Tutorials}, vol.~13, no.~4, pp. 524--540, Nov. 2011.

\bibitem{Madan-1}
R.~Madan, J.~Borran, A.~Sampath, N.~Bhushan, A.~Khandekar, and T.~Ji, ``Cell
  association and interference coordination in heterogeneous {LTE-A} cellular
  networks,'' \emph{IEEE Jour. on Selected Areas in Commun.}, vol.~28, no.~9,
  pp. 1479--1489, Nov. 2010.

\bibitem{Survey_on_HetNet-1}
A.~Damnjanovic, J.~Montojo, Y.~Wei, T.~Ji, T.~Luo, M.~Vajapeyam, T.~Yoo,
  O.~Song, and D.~Malladi, ``A survey on {3GPP} heterogeneous networks,''
  \emph{IEEE Wireless Commun.}, vol.~18, no.~3, pp. 10--21, Jun. 2011.

\bibitem{LTE1}
A.~Khandekar, N.~Bhushan, J.~Tingfang, and V.~Vanghi, ``{LTE-A}dvanced:
  Heterogeneous networks,'' in \emph{Proc. of 2010 European Wireless Conf.},
  pp. 978--982.

\bibitem{3gpp.36.912}
3GPP, ``{{3GPP}; Technical Specification Group Radio Access Network;
  Feasibility study for Further Advancments for {E-UTRA} ({R}elease 9)},''
  {3GPP}, TR {36.912}.

\bibitem{Direct1}
H.~Kim, G.~de~Veciana, X.~Yang, and M.~Venkatachalam, ``Distributed
  $\alpha$-optimal user association and cell load balancing in wireless
  networks,'' \emph{IEEE/ACM Trans. on Networking}, vol.~20, no.~1, pp.
  177--190, Feb. 2012.

\bibitem{Direct2}
K.~Son, S.~Chong, and G.~Veciana, ``Dynamic association for load balancing and
  interference avoidance in multi-cell networks,'' \emph{IEEE Trans. on
  Wireless Commun.}, vol.~8, no.~7, pp. 3566--3576, Jul. 2009.

\bibitem{Andrews-association}
Q.~Ye, B.~Rong, Y.~Chen, M.~Al-Shalash, C.~Caramanis, and J.~Andrews, ``User
  association for load balancing in heterogeneous cellular networks,''
  \emph{IEEE Trans. on Wireless Commun.}, vol.~12, no.~6, pp. 2706--2716, Jun.
  2013.

\bibitem{Andrews2}
Q.~Ye, M.~Al-Shalash, C.~Caramanis, and J.~G. Andrews, ``On/off macrocells and
  load balancing in heterogeneous cellular networks,'' \emph{arXiv preprint
  arXiv:1305.5585}, May. 2013.

\bibitem{Relax1}
S.~Corroy, L.~Falconetti, and R.~Mathar, ``Dynamic cell association for
  downlink sum rate maximization in multi-cell heterogeneous networks,'' in
  \emph{Proc. of 2012 IEEE ICC}, pp. 2457--2461.

\bibitem{MDP1}
E.~Stevens-Navarro, Y.~Lin, and V.~W.~S. Wong, ``An {MDP}-based vertical
  handoff decision algorithm for heterogeneous wireless networks,'' \emph{IEEE
  Trans. on Vehicular Tech.}, vol.~57, no.~2, pp. 1243--1254, Mar. 2008.

\bibitem{MDP2}
S.~Elayoubi, E.~Altman, M.~Haddad, and Z.~Altman, ``A hybrid decision approach
  for the association problem in heterogeneous networks,'' in \emph{Proc. of
  2010 IEEE INFOCOM}, pp. 1--5.

\bibitem{Game1}
E.~Aryafar, A.~Keshavarz-Haddad, M.~Wang, and M.~Chiang, ``{RAT} selection
  games in {H}et{N}ets,'' in \emph{Proc. of 2013 IEEE INFOCOM}, pp. 998--1006.

\bibitem{Game2}
D.~Niyato and E.~Hossain, ``Dynamics of network selection in heterogeneous
  wireless networks: An evolutionary game approach,'' \emph{IEEE Trans. on
  Vehicular Tech.}, vol.~58, no.~4, pp. 2008--2017, May 2009.

\bibitem{GeoSto_Model2}
H.~Dhillon, R.~Ganti, and J.~Andrews, ``Load-aware modeling and analysis of
  heterogeneous cellular networks,'' \emph{IEEE Trans. on Wireless Commun.},
  vol.~12, no.~4, pp. 1666--1677, Apr. 2013.

\bibitem{GeoSto_Model3}
H.-S. Jo, Y.~J. Sang, P.~Xia, and J.~Andrews, ``Heterogeneous cellular networks
  with flexible cell association: A comprehensive downlink {SINR} analysis,''
  \emph{IEEE Trans. on Wireless Commun.}, vol.~11, no.~10, pp. 3484--3495, Oct.
  2012.

\bibitem{GeoSto_Model4}
H.~S.~D. S.~Singh and J.~G. Andrews, ``Offloading in heterogeneous networks:
  Modeling, analysis, and design insights,'' \emph{IEEE Trans. on Wireless
  Commun.}, vol.~12, no.~5, pp. 2484--2497, Dec. 2013.

\bibitem{GeoSto1}
S.~Singh and J.~G. Andrews, ``Joint resource partitioning and offloading in
  heterogeneous cellular networks,'' \emph{IEEE Trans. on Wireless Commun.},
  vol.~13, no.~2, pp. 888--901, Feb. 2014.

\bibitem{cheng2011exploiting}
S.~M. Cheng, S.~Y. Lien, F.~S. Chu, and K.~C. Chen, ``On exploiting cognitive
  radio to mitigate interference in macro/femto heterogeneous networks,''
  \emph{IEEE Trans. on Wireless Commun.}, vol.~18, no.~3, pp. 40--47, Jun.
  2011.

\bibitem{3gpp.36.300}
3GPP, ``Evolved universal terrestrial radio access ({E-UTRA}) and evolved
  universal terrestrial radio access network ({E-UTRAN}),'' {3GPP}, TS {36.300
  v11.7.0}, 2013.

\bibitem{Rappaport}
T.~S. Rappaport, \emph{Wireless Communications: Principles and Practice},
  2nd~ed.\hskip 1em plus 0.5em minus 0.4em\relax Prentice Hall, Inc., 2002.

\bibitem{Boyd_outage}
S.~Kandukuri and S.~Boyd, ``Optimal power control in interference-limited
  fading wireless channels with outage-probability specifications,'' \emph{IEEE
  Trans. on Wireless Commun.}, vol.~1, no.~1, pp. 46--55, Aug. 2002.

\bibitem{Yao_outage}
Y.-D. Yao and A.~Sheikh, ``Outage probability analysis for microcell mobile
  radio systems with cochannel interferers in {R}ician/{R}ayleigh fading
  environment,'' \emph{Electronics Letters}, vol.~26, no.~13, pp. 864--866,
  Jun. 1990.

\bibitem{Hungarian}
H.~W. Kuhn, ``The hungarian method for the assignment problem,'' \emph{Naval
  Research Logistics Quarterly}, vol.~2, pp. 83--397, 1955.

\bibitem{pentico2007assignment}
D.~W. Pentico, ``Assignment problems: A golden anniversary survey,''
  \emph{European Jour. of Operational Research}, vol. 176, no.~2, pp. 774--793,
  Nov. 2007.

\bibitem{boyd2004convex}
S.~P. Boyd and L.~Vandenberghe, \emph{Convex optimization}.\hskip 1em plus
  0.5em minus 0.4em\relax Cambridge university press, 2004.

\bibitem{low1999optimization}
S.~H. Low and D.~E. Lapsley, ``Optimization flow control-'{I}: basic algorithm
  and convergence,'' \emph{IEEE/ACM Trans. on Networking (TON)}, vol.~7, no.~6,
  pp. 861--874, Aug. 1999.

\bibitem{Bertsekas:2009}
D.~P. Bertsekas, \emph{Convex Optimization Theory}.\hskip 1em plus 0.5em minus
  0.4em\relax Athena Scientific, 2009.

\bibitem{Bertsekas:1989}
D.~P. Bertsekas and J.~N. Tsitsiklis, \emph{Parallel and Distributed
  Computation: Numerical Methods}.\hskip 1em plus 0.5em minus 0.4em\relax
  Prentice-Hall, Inc., 1989.

\bibitem{leiserson2001introduction}
C.~E. Leiserson, R.~L. Rivest, C.~Stein, and T.~H. Cormen, \emph{Introduction
  to algorithms}.\hskip 1em plus 0.5em minus 0.4em\relax The MIT press, 2001.

\bibitem{linprogBook}
G.~B. Dantzig, \emph{Linear programming and extensions}.\hskip 1em plus 0.5em
  minus 0.4em\relax Princeton university press, 1998.

\end{thebibliography}

\end{document}